\documentclass[twocolumn]{autart}    % Enable this line and disable the
\usepackage{graphicx}          % Include this line if your
\usepackage{amssymb,amsmath,graphicx,multicol}

\newtheorem{Remark}{Remark}
\newtheorem{Corollary}{Corollary}

\newtheorem{Theorem}{Theorem}
\newtheorem{Proposition}{Proposition}

\newtheorem{Lemma}{Lemma}

\newcommand{\BM}{\begin{matrix}}
\newcommand{\EM}{\end{matrix}}

\newcommand{\ba}{\begin{array}}
\newcommand{\ea}{\end{array}}
\newcommand{\be}{\begin{eqnarray}}
\newcommand{\ee}{\end{eqnarray}}
\newcommand{\EQQ}{\begin{eqnarray*}}
\newcommand{\ENN}{\end{eqnarray*}}

\newcommand{\R}{{\mathbb R}}

\def\tt#1#2{\left[\begin{array}{c}#1\\ #2\end{array}\right]}

\newcommand{\mean}[1]{\mathbb{E}(#1)}
\newcommand{\var}[1]{\mathbb{V}(#1)}

\DeclareMathOperator*{\maximize}{maximize}

\newcommand{\N}{{\mathcal N}}

\begin{document}

\begin{frontmatter}

\title{Maximum entropy properties of discrete-time first-order stable spline kernel \thanksref{footnoteinfo}} % Title, preferably not more
\thanks[footnoteinfo]{This paper was not presented at any IFAC
meeting. Corresponding author T.~Chen. Tel. +46-013284726. }
\vspace{-5mm}
\author[a]{Tianshi Chen}\ead{tschen@isy.liu.se},    % Add the
\author[a]{Tohid Ardeshiri}\ead{tohid@isy.liu.se},               % e-mail address
\author[b]{Francesca P. Carli}\ead{fpc23@cam.ac.uk},
\author[c]{Alessandro Chiuso}\ead{chiuso@dei.unipd.it},
\author[a]{Lennart Ljung}\ead{ljung@isy.liu.se},  % (ead) as shown
\author[c]{Gianluigi Pillonetto}\ead{giapi@dei.unipd.it}

\address[a]{Department of Electrical Engineering, Link\"{o}ping University, Sweden}  % Please supply
\address[b]{Department of Engineering, University of Cambridge, Cambridge, United Kingdom}
\address[c]{Department of Information Engineering, University of Padova, Italy}

                             % keyword list or with the
                                          % help of the Automatica
                                          % keyword wizard

\begin{abstract}                          % Abstract of not more than 200 words.
The first order stable spline (SS-1) kernel is used extensively in
regularized system identification. In particular, the stable spline
estimator models the impulse response as a zero-mean Gaussian
process whose covariance is given by the SS-1 kernel. In this paper,
we discuss the maximum entropy properties of this prior. In
particular, we formulate the exact maximum entropy problem solved by
the SS-1 kernel without Gaussian and uniform sampling assumptions.
Under general sampling schemes, we also explicitly derive  the
special structure underlying the SS-1 kernel (e.g. characterizing
the tridiagonal nature of its inverse), also giving to it a maximum
entropy covariance completion interpretation.  Along the way similar
maximum entropy properties of the Wiener kernel are also given.
\end{abstract}

\end{frontmatter}

\section{Introduction}

A core issue of system identification is the design of model
estimators able to suitably balance structure complexity and
adherence to experimental data. This is also known as the
bias-variance problem in statistical literature.  Traditionally,
this problem is tackled by applying the maximum
likelihood/prediction error method (ML/PEM), see e.g.,
\cite{Ljung:99}, together with model order selection criteria, such
as AIC, BIC and cross validation. Recently, a different method has
been introduced in \cite{PN10a} and further developed in
\cite{PCN11,COL12a,CACLP14}; see also the recent survey
\cite{PDCDL14}. Its key idea is to face the bias-variance problem
via well-designed and tuned regularization. More specifically, the
impulse response $h(t)$ is modeled as a zero-mean Gaussian process $
h(t)\sim\text{GP}(0,k(t,s;\alpha))$, where $k(t,s;\alpha)$ is the
covariance (kernel) function, and $\alpha$ is the hyper-parameter
vector, see e.g., \cite{RasmussenW:06}. The key step is to design a
suitable kernel structure which reflects our prior knowledge on the
system to be identified, e.g., stability. Once $k(t,s;\alpha)$ is
determined, $\alpha$ is tuned by maximizing the marginal likelihood,
and then the posterior mean of $h(t)$ is returned as the impulse
response estimate.

Several kernel structures have been proposed, e.g., the stable
spline (SS) kernel in \cite{PN10a} and the diagonal and correlated
(DC) kernel in \cite{COL12a}, which have shown satisfying
performance via extensive simulated case studies. In view of this,
it seems interesting to investigate how, beyond the empirical
evidence, the use of these regularized approaches can be justified
by theoretical arguments. Different perspectives can be taken, e.g.
deterministic arguments in favor of SS and DC kernels are developed
in \cite{COL12a} while \cite{CCLP14} discusses its link to the
Brownian Bridge process which suggests the first order stable spline
(SS-1) kernel is a natural description for exponentially decaying
impulse responses. In this paper, we will instead work within the
Bayesian context, discussing the maximum entropy (MaxEnt) properties
of the SS-1 kernel.

The MaxEnt approach has been proposed by Jaynes to derive complete
statistical prior distributions from incomplete a priori information
\cite{Jaynes82}. Among all distributions that satisfy some
constraints, e.g. in terms of the value taken by a few expectations,
the MaxEnt criterion chooses the distribution maximizing the
entropy. The justification underlying this choice is that the MaxEnt
distribution, subject to available knowledge, is the one that can be
realized in the greatest number of ways, see also Jaynes'
Concentration Theorem \cite{Jaynes82}.  A preliminary study on the
MaxEnt property of kernels for system identification was developed
in \cite{PN11l}. Working in continuous time (CT), the problem was to
derive the MaxEnt prior using only information on the smoothness and
exponential stability of the impulse response. The arguments in
\cite{PN11l} were however quite involved, mainly due to the
infinite-dimensional nature of the problem and the fact that the
differential entropy rate of a generic CT stochastic process is not
well-defined.  Another recent contribution is \cite{Carli14} where,
under Gaussian and uniform sampling assumptions, it is shown that
the SS-1 kernel matrix can be given a MaxEnt covariance completion
interpretation \cite{Dempster72}, that is then exploited to derive
its special structure (namely that it admits a tridiagonal inverse
with closed form representation as well as factorization).

In this paper, we study the MaxEnt properties of the
\emph{discrete-time} (DT) SS-1 kernel. We first formulate the MaxEnt
problem solved by the DT SS-1 kernel without Gaussian and uniform
sampling assumptions. Then, we extend the result of \cite{Carli14}
and link it to our former result: under general sampling assumption,
we show that the SS-1 kernel matrix is the solution of a maximum
entropy covariance extension problem \cite{Dempster72} with band
constraints. This results in the well-known tridiagonal structure of
the kernel's inverse, which can be also used for efficient numerical
implementations \cite{CL:12},\cite[Section 5]{CCL14}. As a
byproduct, we discuss the MaxEnt properties of the DT Wiener process
and its relation with the tridiagonal structure of the inverse of
its kernel.

\section{MaxEnt property of the Wiener and the SS-1
kernels}\label{sec:tc}

%In this section, we will show how to construct Gaussian processes
%based on top of white Gaussian noise such that they are optimal in
%the sense of MaxEnt subject to constraints among all
%stochastic processes.

Recall that the differential entropy $H(X)$ of a real-valued
continuous random variable $X$ is defined as $
H(X)=-\int_S{p(x)\log{p(x)} \text{d} x}$, where $p(x)$ is the
probability density function of $X$ and $S$ is the support set
of $X$. %In particular, for a Gaussian random variable
%$X\sim \N(\mu,P)$, the differential entropy $H(X) =
%\frac{1}{2}\log\det(P)+\frac{n}{2}\log(2\pi)+\frac{n}{2}$ where
%$X\in\R^n$.

In the sequel, the objects mainly considered are real-valued DT
stochastic processes defined on an ordered index set
$\mathcal{T}=\{t_i| 0\leq t_i<t_{i+1},i=0,1,\cdots,\infty\}$.

A real-valued DT stochastic process $w(t)$ with $t\in\mathcal T$ is
called a white Gaussian noise if $w(t)$ is identically independently
Gaussian distributed with mean $\mean{w(t)}=0$ and variance
$\var{w(t)}=c$.

\subsection{DT Wiener process}

The white Gaussian noise has well-known MaxEnt property. On top of
it, we can construct a more complex Gaussian process with MaxEnt
property which is crucial to derive the MaxEnt property of the SS-1
kernel.

\begin{Lemma}\footnote{All proofs can be found in the Appendix.}
\label{lem:maxent4cgp} Construct a Gaussian process $g(t)$:
\begin{equation}\label{eq:cgp}\begin{aligned} g(t_0)&=0\text{ with
}t_0=0,\\ g(t_k) &= \sum_{i=1}^k w(t_i) \sqrt{t_i-t_{i-1}},
k=1,2,\cdots\end{aligned}\end{equation}  For any $n\in\mathbb N$, it
is the solution to the MaxEnt problem
\begin{equation}\label{eq:maxent4cgp}
\begin{aligned}
\maximize_{h(t)} &\quad H(h(t_1),h(t_2),\cdots,h(t_{n}))\\
\mbox{subject to}\quad &   \var{h(t_{i})-h(t_{i-1})}=
c\left(t_i-t_{i-1}\right)\\& \mean{h(t_i)}=0,  i=1,\cdots, n
\end{aligned}
\end{equation} where it is assumed that $h(t_0)=0$ for $t_0=0$.
\end{Lemma}

The resulting Gaussian process (\ref{eq:cgp}) is actually the DT
Wiener process because it satisfies $g(t_0)=0$, $g(t)$ is Gaussian
distributed with zero mean, and has independent increments with $
g(t_i)-g(t_j)\sim \N\big(0, c(t_i - t_j)\big)$ for $0 \le t_j <
t_i$. It can be verified that the DT Wiener process has zero mean
and covariance (kernel) function:
\begin{equation}\label{eq:DTwiener}
\begin{aligned}  &\text{Wiener:}\quad
K^{\text{Wiener}}(t,s;c)=c\min(t,s), \ t,s\in\mathcal T
\end{aligned}
\end{equation}

%\begin{Lemma}
%\label{lem:Wiener} The discrete-time stochastic process  $g(t)$ on
%$\mathcal{T}$ is  a discrete-time Wiener process if and only if
%$g(t_0)=0$ and
%\begin{equation}\label{eq:wiener}
%g(t_n)=\sum_{i=1}^{n}{h(t_i)}\sqrt{t_i-t_{i-1}}, \  n\geq1
%\end{equation}
%where $h(t)$ is the zero-mean Gaussian white noise.% with variance $\lambda$. Then, $g(t)$ is Gaussian process and covariance of $g(t)$ is given by
%%\begin{equation}
%%\mathbb{V}[g(t_i),g(t_j)]=\lambda \min(t_i,t_j)\  \text{ for } \ t_i,t_j \in \mathcal{T}
%%\end{equation}
%%=========
%\end{Lemma}

\subsection{The first order SS kernel}\label{sec:tc}

Based on Lemma \ref{lem:maxent4cgp}, we can derive the MaxEnt
property for the SS-1 kernel:
\begin{equation}\label{eq:SS-1}
\begin{aligned}  &\text{SS-1:}\quad
K^{\text{SS-1}}(t,s;\alpha)=c\min(e^{-\beta t},e^{-\beta
s}),\\&\qquad\qquad\qquad \alpha = [c\ \beta]^T, c\geq0,\beta>0, \
t,s\in\mathcal T
\end{aligned}
\end{equation}
It is also introduced independently in a deterministic argument in
\cite{COL12a} and called the tuned correlated (TC) kernel. It is
fair to call (\ref{eq:SS-1}) the SS-1 kernel here, since the
``stable'' time transformation involved in deriving the SS-1 kernel
plays a key role in the following theorem.

\begin{Theorem}\label{thm:maxent4tc} Let $w(\cdot)$ be a white Gaussian noise with mean zero
and variance $c$. Then the stochastic process
\begin{equation}
\begin{aligned}\label{eq:sol2maxent4tc} h^o(t_k) &= \sum_{i=k}^{n-1} w(e^{-\beta
t_i}) \sqrt{e^{-\beta t_i} - e^{-\beta t_{i+1}}},\\& k=0,\cdots,n-1,
h^o(t_n) =0 \text{ with } t_n=\infty
\end{aligned}
\end{equation}  is a Gaussian process with zero mean and
the SS-1 kernel (\ref{eq:SS-1}) as its covariance function, and for
any $n\in\mathbb N$, it is the solution to the MaxEnt problem
\begin{align}
\maximize_{h(t)} &\quad H(h(t_0),h(t_1),\cdots,h(t_{n-1}))\nonumber\\
\mbox{subject to}\quad &  \var{h(t_{i+1})-h(t_{i})}= c\left(
e^{-\beta t_i}-e^{-\beta t_{i+1}}\right)\nonumber\\ &
\mean{h(t_i)}=0, i=0,\cdots, {n-1}\label{eq:maxent4tc}
\end{align}
where it is assumed that $h(t_n)=0$ with $t_n=\infty$.
\end{Theorem}
\begin{Remark} In the optimization criteria (\ref{eq:maxent4cgp}) and
(\ref{eq:maxent4tc}), if we divide the entropy of the sequence of
the stochastic process by $n$ and let $n$ go to $\infty$, then the
limit (if exists) becomes the differential entropy rate of the
stochastic process \cite{CT12}. However, the limit does not exist
for Gaussian processes (\ref{eq:cgp}) and (\ref{eq:sol2maxent4tc}),
which is the reason why the entropy of a sequence of stochastic
processes is used here instead.
\end{Remark}

\section{Special structure of Wiener and SS-1 kernels and their MaxEnt interpretation}

%In this section, we first show that the inverse of the kernel matrix
%of the Wiener kernel and the SS-1 kernel are tridiagonal and we
%then show that this special structure can be given a MaxEnt
%interpretation.

%\subsection{Special structure}

In what follows, we let $c=1$ and consider kernel matrix $P$ with
dimension $n\geq3$ defined as \begin{align} P_{i,j} =
K(t_i,t_j;\alpha),\ i,j=1,\cdots,n,\ t_i,t_j\in\mathcal T
\end{align}
where $P_{i,j}$ denotes the $(i,j)$th element of the matrix $P$ and
$K$ is either the Wiener kernel (\ref{eq:DTwiener}) or the SS-1
kernel (\ref{eq:SS-1}). We find that $P$ has some special structure,
e.g., its inverse is tridiagonal and its square root has closed-form
expression. These special structure can be used to improve the
stability and efficiency of the implementation solving the marginal
likelihood maximization, see e.g., \cite[Remark 4.2]{CL:12},
\cite[Section 5]{CCL14}.

\begin{Proposition}\label{prop:tc_structure} Consider the Wiener kernel (\ref{eq:DTwiener}) and the SS-1 kernel
(\ref{eq:SS-1}). Then the following results hold:
\begin{enumerate}
\item[(a)] for the Wiener kernel, $\det (P^{\text{Wiener}}) = t_1\Pi_{k=1}^{n-1}
(t_{k+1}-t_{k})$ and $(P^{\text{Wiener}})_{i,j}^{-1}$ is equal to
\begin{align*} \left\{\begin{array}{cc}
                                                                                                                     \frac{t_2}{t_1(t_2-t_1)},
                                                                                                                     &
                                                                                                                     i=j=1,\\
                                                                                                                     \frac{t_{i+1}-t_{i-1}}{(t_{i+1}-t_{i})(t_{i}-t_{i-1})},
                                                                                                                     &
                                                                                                                     i=j=2,\cdots,n-1,\\
                                                                                                                     \frac{1}{t_{n}-t_{n-1}},
                                                                                                                     &
                                                                                                                     i=j=n,\\
                                                                                                                     0,
                                                                                                                     &
                                                                                                                     |i-j|>1\\   -\frac{1}{\max(t_i,t_j)-\min(t_i,t_j)}, &  \emph{otherwise},
                                                                                                                   \end{array}
\right.
\end{align*}

\item[(b)] for the SS-1 kernel, $\det (P^{\text{SS-1}}) = e^{-\beta t_n}\Pi_{k=1}^{n-1}
(e^{-\beta t_k }-e^{-\beta t_{k+1} })$ and
$(P^{\text{SS-1}})_{i,j}^{-1}$ is equal to
\begin{align*} \left\{\begin{array}{cc}
                                                                                                                     \frac{1}{e^{-\beta t_1}-e^{-\beta t_2}},
                                                                                                                     &
                                                                                                                     i=j=1,\\
                                                                                                                     \frac{e^{-\beta t_{i-1}}-e^{-\beta t_{i+1}}}{(e^{-\beta t_{i-1}}-e^{-\beta t_{i}})(e^{-\beta t_{i}}-e^{-\beta t_{i+1}})},
                                                                                                                     &
                                                                                                                     i=j=2,\cdots,n-1,\\
                                                                                                                     \frac{e^{-\beta(t_{n-1}-t_n)}}{e^{-\beta t_{n-1}}-e^{-\beta t_{n}}},
                                                                                                                     &
                                                                                                                     i=j=n,\\
                                                                                                                     0,
                                                                                                                     &
                                                                                                                     |i-j|>1\\   -\frac{1}{e^{-\beta\min\{t_i,t_j\}}-e^{-\beta \max\{t_i,t_j\}}}, &  \emph{otherwise},
                                                                                                                   \end{array}
\right.
\end{align*}
\end{enumerate}
\end{Proposition}

\begin{Corollary}\label{coro:tc_structure} Consider the Wiener kernel (\ref{eq:DTwiener}) and the SS-1 kernel
(\ref{eq:SS-1}). Then the following results hold:
\begin{enumerate}
\item[(a)] for the Wiener kernel,
\begin{align}\label{eq:Wienerdecomp}
(P^{\text{Wiener}})^{-1} = W^TW\end{align} where $W$ is upper
bidiagonal with
\begin{align*}
&W(i,i) = -\frac{t_{i+1}}{t_i} W(i,i+1) =
\sqrt{\frac{t_{i+1}}{t_i}\frac{1}{t_{i+1}-t_i}},\\&\qquad
i=1,\cdots,n-1,\ W(n,n) = \sqrt{1/t_n}
%W(i,i+1)=-\sqrt{\frac{t_i}{t_{i+1}}\frac{1}{t_{i+1}-t_i}}, \\
%&\qquad\qquad
\end{align*}

\item[(b)] for the SS-1 kernel, \begin{align}\label{eq:tcdecomp}(P^{\text{SS-1}})^{-1} = S^TS\end{align} where
$S$ is upper bidiagonal with
\begin{align*}
&S(i,i) = -S(i,i+1)=\frac{1}{\sqrt{e^{-\beta t_i}-e^{-\beta t_{i+1}}}}, \\
& i=1,\cdots,n-1,\ S(n,n) =
\sqrt{\frac{e^{\beta(t_n-t_{n-1})}-1}{e^{-\beta t_{n-1}}-e^{-\beta
t_{n}}}}
\end{align*}
\end{enumerate}
\end{Corollary}

\begin{Remark}\label{rmk:tc_structure}
From (\ref{eq:Wienerdecomp}) and (\ref{eq:tcdecomp}), decomposing
$P=UU^T$ for upper triangular $U$ has closed form expression. For
the Wiener kernel, $U=W^{-1}$ with $U_{i,j}=(W_{i,i})^{-1}t_i/t_j$
for $i\geq j$, $i,j=1,\cdots,n$. For the SS-1 kernel, $U=S^{-1}$
with $U_{i,j}=(S_{i,i})^{-1}$ for $i\geq j$, $i,j=1,\cdots,n$.
\end{Remark}

\begin{Remark}\label{rmk:tc_structure}
Recall from e.g., \cite{Dempster72} that if $X\sim\N(0,P)$ with
$P^{-1}_{i,j}=0$, then $X_i$ and $X_j$ are conditionally independent
given $X_k$ with $k\neq i,j$ where $X_k$ is the $k$th element of
$X$. This means that the Wiener and SS-1 kernels correspond to
sparse representation, see e.g., \cite{Dempster72} for details and
also the proof of Corollary \ref{coro:tc_structure}.
\end{Remark}

\subsection{MaxEnt covariance completion}

The fact that the kernel matrices of the Wiener and SS-1 kernels
have tridiagonal inverse can be given a MaxEnt covariance completion
interpretation.

Recall that a real symmetric matrix $A$ with dimension $n> m+1$ is
called an $m-$\emph{band} matrix if $A_{i,j}=0$ for $|i-j|>m$, and
the matrix $M$ is called an extension of $A$ if $M_{i,j}=A_{i,j}$
for $|i-j|\leq m$. Moreover, $M$ is called a positive extension of
$A$ if $M$ is positive definite. A positive extension $M$ of the
$m-$band matrix $A$ is called a band-extension of $A$ if $M^{-1}$ is
an $m-$band matrix.

\begin{Theorem}\label{thm:covext} Define $A\in\R^{n\times n}$ as follows:
\begin{align}\label{eq:A_wienere&ss-1} A_{i,j}&=\left\{\begin{array}{cc}
                                                   P^{\text{Wiener}}_{i,j}\ (\text{resp.}\ P^{\text{SS-1}}_{i,j}),  & |i-j|\leq1 \\
                                                   0 & |i-j|>1
                                                 \end{array}
 \right.
\end{align}
Then $P^{\text{Wiener}}$ (resp. $P^{\text{SS-1}}$) is the unique
band extension of $A$, and the Gaussian random vector with zero mean
and covariance $P^{\text{Wiener}}$ (resp. $P^{\text{SS-1}}$) is the
unique solution to the MaxEnt covariance completion problem
\begin{equation}\label{eq:maxent4covext4wiener&ss-1}
\begin{aligned}
\maximize_{P} &\quad H(X)\\
\mbox{subject to}\quad & \text{P is any positive extension of $A$}
\end{aligned}
\end{equation} where $X$ is a zero mean random vector with covariance matrix
$P$.
\end{Theorem}

\begin{Remark}
To our best knowledge, for the Wiener kernel (\ref{eq:DTwiener}) the
special structure and its MaxEnt interpretation has not been pointed
out before. For the SS-1 kernel (\ref{eq:SS-1}), the result under
the uniform sampling assumption is given in \cite{Carli14}  and thus
is a special case of this paper.
\end{Remark}

\section{Conclusion}
We have shown that a zero mean Gaussian process with the first-order
stable spline kernel solves a maximum entropy problem with the
constraint that the variance of neighboring impulse response
coefficients at $t_i<t_{i+1}$ is proportional to $e^{-\beta
t_i}-e^{-\beta t_{i+1}}$, which decays to zero ultimately. Its
kernel matrix (also true for the Wiener kernel) solves a maximum
entropy covariance completion problem and has special structure,
e.g., its inverse is tridiagonal, under general sampling
assumptions. Finally, one may wonder if the other kernels, e.g., the
diagonal correlated kernel, can be given similar maximum entropy
interpretation. The answer is more involved and will be discussed
separately, see e.g., \cite{CCL14}.

\bibliographystyle{unsrt}        % Include this if you use bibtex
\bibliography{../ref}          % and a bib file to produce the
                                 % bibliography (preferred). The
                                 % correct style is generated by
                                 % Elsevier at the time of printing.

%\begin{thebibliography}{99}     % Otherwise use the
                                 % thebibliography environment.
                                 % Insert the full references here.
                                 % See a recent issue of Automatica
                                 % for the style.
%  \bibitem[Heritage, 1992]{Heritage:92}
%     (1992) {\it The American Heritage.
%     Dictionary of the American Language.}
%     Houghton Mifflin Company.
%  \bibitem[Able, 1956]{Abl:56}
%     B.~C.~Able (1956). Nucleic acid content of macroscope.
%     {\it Nature 2}, 7--9.
%  \bibitem[Able {\em et al.}, 1954]{AbTaRu:54}
%     B.~C. Able, R.~A. Tagg, and M.~Rush (1954).
%     Enzyme-catalyzed cellular transanimations.
%     In A.~F.~Round, editor,
%     {\it Advances in Enzymology Vol. 2} (125--247).
%     New York, Academic Press.
%  \bibitem[R.~Keohane, 1958]{Keo:58}
%     R.~Keohane (1958).
%     {\it Power and Interdependence:
%     World Politics in Transition.}
%     Boston, Little, Brown \& Co.
%  \bibitem[Powers, 1985]{Pow:85}
%     T.~Powers (1985).
%     Is there a way out?
%     {\it Harpers, June 1985}, 35--47.

%\end{thebibliography}

\section*{Appendix}
\renewcommand{\thesubsection}{\Alph{subsection}}

%\subsection{Proof of Lemma \ref{lem:maxent4wgn}}    % Each appendix must have a short title.
%The proof is a straightforward adaption of Burg's MaxEnt
%theorem in \cite[Theorem 12.6.1]{CT12} and thus is omitted.

\subsection*{Proof of Lemma \ref{lem:maxent4cgp}}

%The proof is straightforward but is still given below for
%completeness.
First, we recall the well-known MaxEnt property of white Gaussian
noise.

\begin{Lemma} \cite[Burg's MaxEnt Theorem, page 417]{CT12}
\label{lem:maxent4wgn_reduced} Consider the white Gaussian noise
$w(t)$. For any $n\in\mathbb N$, it is the solution to the MaxEnt
problem:
\begin{align}\label{eq:maxent4wgn}
\maximize_{r(t)} &\quad
H(r(t_0),r(t_1),\cdots,r(t_{n-1}))\\\nonumber \mbox{subject to}\quad
& \mean{r(t_i)}=0, \var{r(t_i)}=c, i=0,...,n-1
\end{align}
\end{Lemma}

Then from (\ref{eq:maxent4cgp}), define $v(t_i) =
\frac{h(t_{i})-h(t_{i-1})}{\sqrt{t_i-t_{i-1}}}$, $i=1,\cdots,n$. We
have $\mean{v(t_i)}=0,\var{v(t_i)}=c$, $i=1,\cdots,n$, and
\begin{align}\label{eq:intermediate} h(t_k) = \sum_{i=1}^k v(t_i)
\sqrt{t_i-t_{i-1}}, k=1,2,\cdots,n
\end{align}
Now let $L=[h(t_1)\ h(t_2)\ \cdots\ h(t_n)]^T$, $V=[v(t_1)\
v(t_2)\\\
\cdots\ v(t_n)]^T$, and $B$ be a lower-triangular matrix with
$B_{i,j}=\sqrt{t_j-t_{j-1}}$ for $i\geq j$. Then  we have $L=BV$.
%\begin{align}\label{eq:matrix} \left[
%  \begin{array}{c}
%    h(t_1) \\
%    h(t_2) \\
%    \vdots \\
%    h(t_n) \\
%  \end{array}
%\right] = \begin{array}{c}
%            \underbrace{\left[
%            \begin{array}{cccc}
%              b(t_1,t_0) & 0 & \cdots & 0 \\
%              b(t_1,t_0) & b(t_2,t_1) & \cdots & 0 \\
%              \vdots & \vdots & \cdots & 0 \\
%              b(t_1,t_0) & b(t_2,t_1) & \cdots & b(t_n,t_{n-1}) \\
%            \end{array}
%          \right]} \\
%            M
%          \end{array}
%\left[
%  \begin{array}{c}
%    v(t_1) \\
%    v(t_2) \\
%    \vdots \\
%    v(t_n) \\
%  \end{array}
%\right]
%\end{align} where $b(t_i,t_{i-1})= t_i-t_{i-1}>0$ for $i=1,\cdots,n$ and the matrix $M$ in
%(\ref{eq:matrix}) is lower-triangular
Apparently, $B$ is nonsingular in that all main diagonal elements
are strictly positive. Further noting the property (see e.g.,
\cite[Corollary to Theorem 8.6.4]{CT12}) that
$H(L)=H(V)+\log\det(B)$
%\begin{align*}
%&H(h(t_1),h(t_2),\cdots,h(t_{n}))=  H(v(t_1),v(t_2),\cdots,v(t_{n}))
%\\&\qquad\qquad+ \log\det (M)
%\end{align*}
yields that the MaxEnt problem (\ref{eq:maxent4cgp}) is equivalent
to
\begin{equation}\label{eq:maxent4cgp_2}
\begin{aligned}
\maximize_{v(t)} &\quad H(v(t_1),v(t_2),\cdots,v(t_{n})) + \log\det (B)\\
\mbox{subject to}&\quad  \mean{v(t_i)}=0, \var{v(t_{i})}= c,
i=1,\cdots, n
\end{aligned}
\end{equation} Since the matrix $B$ is independent of $v(t)$ or $h(t)$, the maximum
entropy problem (\ref{eq:maxent4cgp_2}) is further equivalent to
(\ref{eq:maxent4wgn}). As a result, the optimal $v(t)$ to
(\ref{eq:maxent4cgp_2}) is the  white Gaussian noise $w(t)$.
Finally, comparing (\ref{eq:intermediate}) with (\ref{eq:cgp})
yields that the constructed Gaussian process $g(t)$ in
(\ref{eq:cgp}) is indeed the optimal solution to
(\ref{eq:maxent4cgp}).

%\subsection{Proof of Lemma \ref{lem:Wiener}}
%
%First, we prove the necessary part. That is, we show if $g(t)$ is a
%discrete-time Wiener process, it can be expressed in the form of
%(\ref{eq:wiener}). Let $w(t_i)\triangleq
%\frac{g(t_i)-g(t_{i-1})}{\sqrt{t_i - t_{i-1}}}$ for
%$i\in\mathbb{N}$. Since $ g(t_i)-g(t_{i-1})\thicksim \N\big(0,
%\lambda(t_i - t_{i-1})\big)$ we have
%\begin{equation}
%w(t_i)\thicksim \N\big(0, \lambda\big).
%\end{equation}
%Also, since $g(t)$ has independent increments it follows that $w(t)$
%is a discrete-time zero-mean Gaussian white noise. Also,
%from the definition of $w(t)$ we have
%\small{\begin{equation}\nonumber
%g(t_n)={w(t_n)}\sqrt{t_n-t_{n-1}}+g(t_{n-1})=\sum_{i=1}^{n}{w(t_i)}\sqrt{t_i-t_{i-1}}.
%\end{equation}}\normalsize
%Now we prove the sufficient part, i.e., the stochastic process
%(\ref{eq:wiener}) is a discrete-time Wiener process. Since Gaussian
%processes are closed under linear operations \cite{RasmussenW:06},
%$g(t)$ is a Gaussian process. Also
%$\mean[g(t_n)]=\mean[\sum_{i=1}^{n}{h(t_i)}\sqrt{t_i-t_{i-1}}]=0$.
%Furthermore let $0 \le t_j < t_i$,
%\begin{align*}
%&\text{Var}[g(t_i)-g(t_j)]=\mean[\big(\sum_{r={j+1}}^{i}{h(t_r)\sqrt{t_r-t_{r-1}}}\big)^2]\\
%&=\sum_{r={j+1}}^{i}{\lambda ({t_r-t_{r-1}})}=\lambda (t_i-t_j)
%\end{align*}
%and . Therefore, $ g(t_i)-g(t_j)\thicksim \N\big(0, \lambda(t_i -
%t_j)\big)$  for $0 \le t_j < t_i$. Since $h(t)$ is a Gaussian white
%noise the increments of $g(t)$ are independent and the proof
%follows.

\subsection*{Proof of Theorem \ref{thm:maxent4tc}}
We first introduce a time transformation, and define \begin{align}
\label{eq:timetransform}\tau_i&=e^{-\beta
t_{n-i}},\\\label{eq:timetransform2} f(\tau_i) &= h(-\log
(\tau_i)/\beta),\ i=0,\cdots,n.\end{align}
%Then the MaxEnt problem (\ref{eq:maxent4tc}) is equivalent
%to
%\begin{equation}\nonumber
%\begin{aligned}
%\maximize_{h} &\quad H(h(-\log (\tau_n/\beta)),\cdots,h(-\log (\tau_1/\beta)))\\
%\mbox{subject to} &\quad \var{h(-\log (\tau_{n-i-1}/\beta))-h(-\log
%(\tau_{n-i}/\beta))}\\&\qquad\qquad= c\left(
%\tau_{n-i}-\tau_{n-i-1}\right)\\& \mean{h(-\log
%(\tau_{n-i}/\beta))}=0, i=0,\cdots, {n-1}
%\end{aligned}
%\end{equation} where $h(-\log (\tau_0/\beta))=0$ with
%$\tau_0=0$. Now define
%\begin{align}\label{eq:timetransform2} f(\tau_i) =
%h(-\log (\tau_i/\beta)), i=0,\cdots,n
%\end{align}
Then the MaxEnt problem (\ref{eq:maxent4tc}) is equivalent
to\begin{equation}\label{eq:maxent4tc3}
\begin{aligned}
\maximize_{f(\tau)} &\quad H(f(\tau_1),f(\tau_2),\cdots,f(\tau_n))\\
\mbox{subject to}\quad & \var{f(\tau_{i})-f(\tau_{i-1})}= c\left(
\tau_{i}-\tau_{i-1}\right)\\& \mean{f(\tau_i)}=0, i=1,\cdots, n
\end{aligned}
\end{equation} where it is assumed that $f(\tau_0)=0$ with
$\tau_0=0$. By Lemma \ref{lem:maxent4cgp}, the optimal solution to
(\ref{eq:maxent4tc3}) is the Gaussian process $g(\tau)$ defined as
follows: \begin{equation}\begin{aligned}\label{eq:cgptc}
g(\tau_0)&=0\text{ with }\tau_0=0,\\ g(\tau_k) &= \sum_{i=1}^k
w(\tau_i) \sqrt{\tau_i-\tau_{i-1}}, k=1,2,\cdots\end{aligned}
\end{equation}
 where $w(\tau)$ is the white Gaussian noise
defined on $\{\tau_0,\tau_1,\cdots\}$.  Finally, noting
(\ref{eq:timetransform2}) and (\ref{eq:timetransform}) yields that
the optimal solution to (\ref{eq:maxent4tc}) is
(\ref{eq:sol2maxent4tc}). Apparently, (\ref{eq:sol2maxent4tc}) is a
Gaussian process with zero mean and the SS-1 kernel as its
covariance function. This completes the proof.

\subsection*{Proof of Proposition \ref{prop:tc_structure}}

For the proof of the results hereafter, we only give the proof for
the SS-1 kernel and that for the Wiener kernel can be derived in the
same way and thus is omitted.

From (\ref{eq:sol2maxent4tc}), define $x=[x_1,\cdots,x_n]^T$ with
$x_k=h^o(t_{k})-h(t_{k-1})$, $k=1,\cdots,n$. Then we have
$x\sim\N(0,Q)$ where $Q$ is a diagonal matrix with
$Q_{i,i}=e^{-\beta t_{i-1}}-e^{-\beta t_{i}}$, $i=1,\cdots,n$.
Moreover, $ (P^{\text{SS-1}})^{-1} = V^{-T} Q^{-1}V$ where $V$ is an
upper bidiagonal matrix with all main diagonal elements equal to
$-1$ and the first upper off-diagonal elements equal to $1$.
Apparently, $(P^{\text{SS-1}})^{-1}$ takes the form in part b),
which completes the proof.

\subsection*{Proof of Corollary \ref{coro:tc_structure} }

By completing the squares, $ \theta^T(P^{\text{SS-1}})^{-1}\theta =
\sum_{k=1}^{n-1}S_{k,k}^2\\ (\theta_k-\theta_{k+1})^2 + S_{n,n}^2
\theta_n^2 $ where $\theta\in\R^n$ and $\theta_k$ is the $k$th
element of $\theta$. Then (\ref{eq:tcdecomp}) follows immediately.

\subsection*{Proof of Theorem \ref{thm:covext}}

%To prove this theorem, we recall two lemmas from band matrix
%extension problems: the first one is a result of \cite[Theorem 1.1,
%page 892, Corollary 1.2, page 897]{GGK93}, and the second one is a
%result of \cite[Theorem 2.1, page 898, Theorem 2.2, page 899,
%Corollary 1.5, page 945]{GGK93}.

We first recall a lemma from band matrix
extension problems, that is a
result of \cite[Theorem 2.1, page 898, Theorem 2.2, page 899,
Corollary 1.5, page 945]{GGK93}.

%\begin{Lemma}\cite{GGK93}\label{lem:covext_1stepext} Consider a real symmetric matrix
%\begin{align*} A(z) = \left(
%                                        \begin{array}{ccccc}
%                                          A_{1,1} & A_{1,2} & \cdots & A_{1,n-1} & z \\
%                                          A_{2,1} & A_{2,2} & \cdots & A_{2,n-1} & A_{2,n} \\
%                                          \vdots & \vdots & \vdots & \vdots & \vdots \\
%                                          A_{n-1,1} & A_{n-1,2} & \cdots & A_{n-1,n-1} & A_{n-1,n} \\
%                                          z & A_{n,2} & \cdots & A_{n,n-1} & A_{n,n} \\
%                                        \end{array}
%                                      \right)
%\end{align*} with all elements given and fixed except
%$A_{1,n}=A_{n,1}=z$. If the following submatrices are nonsingular
%\begin{align}\label{eq:matricetest}
%L= [A]_{1}^{n-1},Q= [A]_{2}^{n-1}, R= [A]_{2}^{n},
%\end{align} where $[A]_{s}^l$ with $s\leq l$ denotes the
%submatrix of $A$ from the $s$th row (resp. column) to the $l$th row
%(resp. column), then the central one-step extension $A(z_0)$ with
%\begin{equation}\begin{aligned} \label{eq:z0}
%&z_0=-\frac{1}{y_1}\sum_{j=2}^{n-1} a_{nj}y_j\\& \left[
%  \begin{array}{ccccc}
%    y_1 &
%    y_2 &
%    \cdots &
%    y_{n-1} &
%  \end{array}
%\right]^T = L^{-1} \left[
%  \begin{array}{ccccc}
%    1 &
%    0 &
%    \cdots &
%    0 &
%  \end{array}
%\right]^T
%\end{aligned}\end{equation} is the unique one-step extension of $A(0)$ with the
%property that $(A(z_0)^{-1})_{1,n}=(A(z_0)^{-1})_{n,1}=0$.
%\end{Lemma}

\begin{Lemma}\cite{GGK93}\label{lemma:covext} Assume that $A$ is an $m$-band matrix with dimension
$n>m+1$ and that the submatrices $[A]_i^{m+i}$, $i=1,\cdots,n-m$,
are positive definite, where $[A]_{s}^l$ with $s\leq l$ denotes the
submatrix of $A$ from the $s$th row (resp. column) to the $l$th row
(resp. column). Then we have: \begin{enumerate}
%\item[(a)] There exists a unique positive extension $M$ of $A$ with
%the property that for all $m+2\leq l\leq n$ and $1\leq s \leq l -
%m-1$ , $[M]_s^l$ is the central one-step extension of the
%corresponding $(l-s-1)$-band matrix.

\item[(a)] $M$ is the unique band extension of $A$.

\item[(b)] The Gaussian random vector with zero mean and covariance matrix
$M$ is the unique solution to the  MaxEnt problem
\begin{equation}\label{eq:maxent4covext}
\begin{aligned}
\maximize_{X} &\quad H(X)\\
\mbox{subject to}\quad & \text{P is any positive extension of $A$}
\end{aligned}
\end{equation} where $X$ is a zero mean random vector with covariance matrix $P$.
\end{enumerate}
\end{Lemma}

Apparently, $A$ in (\ref{eq:A_wienere&ss-1}) is a $1-$ band matrix
and $[A]_i^{i+1}$, $i=1,\cdots,n-1$, are positive definite. This
means that the results of Lemma \ref{lemma:covext} hold for $A$ in
(\ref{eq:A_wienere&ss-1}) and the remaining task is to show
$M=P^{\text{SS-1}}$, i.e., the optimal solution $P^{\text{Opt}}$ of
(\ref{eq:maxent4covext}) is $P^{\text{Opt}}=P^{\text{SS-1}}$. This
task can be accomplished by noting the relation between the problems
(\ref{eq:maxent4covext}) and (\ref{eq:maxent4tc}). Note that the
Gaussian process (\ref{eq:sol2maxent4tc}) solves the problem
(\ref{eq:maxent4tc}) and has the SS-1 kernel as its covariance
function. Assume $Z\sim\N(0,P)$. Then for $n\geq3$, the covariance
matrix $P^{\text{SS-1}}$ is the optimal solution to
\begin{align}\nonumber
\maximize_{P} &\quad H(Z)\\\nonumber \mbox{subject to}\quad &
P_{i,i}+P_{i+1,i+1}-2P_{i,i+1}\\\nonumber&=e^{-\beta t_i} -
e^{-\beta t_{i+1}}, i=1,\cdots,n-1\\& \text{$P$ is positive
definite}\label{eq:maxent4covext2}
\end{align}
Also note that the constraint in (\ref{eq:maxent4covext}) is a
subset of the constraint in (\ref{eq:maxent4covext2}), hence $
H(X)\leq H(Z)$ with $X\sim\N(0,P^{\text{Opt}})$ and
$Z\sim\N(0,P^{\text{SS-1}})$. Finally, noting that $P^{\text{SS-1}}$
is a positive extension of $A$ and the uniqueness of
$P^{\text{Opt}}$ yields $P^{\text{Opt}}=P^{\text{SS-1}}$. This
completes the proof.

\end{document}